\documentclass[twocolumn,preprintnumbers,amsmath,amssymb]{revtex4}

\usepackage{array}
\usepackage{graphicx}
\usepackage{tabularx}
\usepackage{longtable}
\usepackage{dcolumn} 
\usepackage{bm}

\begin{document}

\title{Large thermal Hall coefficient in bismuth
}

\author{W. Kobayashi$^{*,1,2}$, Y. Koizumi$^3$, and Y. Moritomo$^1$}
\affiliation
{$^1$Graduate School of Pure and Applied Sciences, University of Tsukuba, 
Ibaraki 305-8571, Japan}
\affiliation
{$^2$PRESTO, Japan Science and Technology Agency, Saitama 332-0012, Japan}
\affiliation
{$^3$Department of Physics, Waseda University, Tokyo 169-8555, Japan}
\email{kobayashi.wataru.gf@u.tsukuba.ac.jp}

\date{\today}

\begin{abstract}

We present a systematical study of thermal Hall effect on a bismuth single crystal by 
measuring resistivity, Hall coefficient, and thermal conductivity under magnetic field, 
which shows a large thermal Hall coefficient comparable to the largest one in 
a semiconductor HgSe. We discuss that this is mainly due to a large mobility 
and a low thermal conductivity comparing theoretical calculations, 
which will give a route for controlling heat current in electronic devices.

\end{abstract}

\maketitle

Thermal Hall effect (Righi-Leduc effect) is the thermal analog of the Hall effect. 
A temperature gradient $\frac{dT_x}{dx}$ introduced along $x$ direction yields a 
temperature gradient $\frac{dT_y}{dy}$ along $y$ direction 
in the presence of the magnetic field $B_z$ along $z$ direction. 
A ratio of the temperature gradients is defined as 
a thermal Hall coefficient $R_{\rm TH}$, 
\begin{equation}
R_{\rm TH} \equiv \frac{1}{B_z}\frac{\frac{dT_y}{dy}}{\frac{dT_x}{dx}}=\sigma R_{\rm H}=\mu _{\rm H},  
\end{equation}
which shows a measure of oriented heat current per magnetic field. 
The definition of $R_{\rm TH}$ differs from that of the 
Hall coefficient $R_{\rm H}$. 
$R_{\rm TH}$ and $R_{\rm H}$ are inseparably related, 
since they are both governed by the Fermi surface topology. 
The relationship between $R_{\rm TH}$ and $R_{\rm H}$ ($\sigma$: electrical conductivity) 
shown in Eq. (1) was derived by Bridgman 
\cite{bridgman} by considering heat and electric transport in terms of 
reversible processes in thermodynamics, and qualitatively explains the thermal Hall effect in 
conventional metals \cite{stephan,fletcher}. 
The thermal Hall effect of semiconductors has been studied until 1960s 
\cite{HgSe,mette}. 
As a result, it is known that $R_{\rm TH}$ of the semiconductors 
satisfies a relationship \cite{putley}, 
\begin{equation}
R_{\rm TH}\cong \frac{\kappa _{\rm el}}{\kappa _{\rm tot}}\mu _{\rm H} 
\end{equation}
that takes into account heat conduction along the longitudinal direction 
through lattices besides electrons \cite{note}, 
where $\kappa _{\rm el}$, $\kappa _{\rm tot}$($=\kappa _{\rm ph} +\kappa _{\rm el}$, 
$\kappa _{\rm ph}$: lattice thermal conductivity), 
and $\mu _{\rm H}$ represent 
electron thermal conductivity, total thermal conductivity, and Hall mobility, respectively. 
Recent developments of this research field have revealed phonon and 
magnon thermal Hall effects, in which carriers of heat currents along the $y$ direction are not 
electrons but phonons and magnons \cite{strohm}. These extreme effects have a potential for controlling 
heat current in electronic devices as well as a thermal rectifier 
\cite{chang,casati,BLi,kobayashi}.

Bismuth is "an old material", which has been extensively studied 
owing to its unique physical properties such as 
large positive magnetoresistance, huge orbital diamagnetism, and 
peculiar magnetic-field dependent thermoelectric properties \cite{edelman}. 
Recent increasing interests in Dirac 
fermions and strong spin-orbit interaction in a heavy element such 
as Bi, Pt, or Au motivate many researchers to study this material. 
Since the bismuth exhibits large $\mu _{\rm H}$, relatively large 
$\kappa _{\rm el}$, and low lattice thermal conductivity $\kappa _{\rm ph}$ 
due to both a semimetallic state and the heavy mass of Bi element, 
a large $R_{\rm TH}$ 
is expected through Eqs. (1) and (2). Though $R_{\rm TH}$ of bismuth 
has been already reported in 1910s \cite{smith}, there exists the data only at 
around room temperature and low magnetic field. 
Thus, in this letter, we report thermal Hall effect on a bismuth single crystal up to 
3 T and in the temperature range of 75 to 300 K, and first analyze temperature 
dependent $R_{\rm TH}$ using these equations. 
We also discuss a possible way to further 
increase $R_{\rm TH}$.

Resistivity, Hall coefficient and thermal conductivity measurements of 
a high quality single crystal (Bi 99.999\%) with a 
dimension of $1.15\times 2.35\times 5.15$ mm$^3$ were performed. 
For the measurements, electric current $i_x$ and heat current $j_x$ were 
applied along the bisectrix axis under the magnetic field introduced parallel 
to the trigonal axis of the bismuth single crystal. 
Induced Hall voltage $V_y$ and thermal Hall temperature 
gradient $\frac{\bigtriangleup T_y}{\bigtriangleup y}$ were detected along the binary axis. 
The resistivity and Hall coefficient were measured by a conventional four probe method. 

The thermal conductivity and the thermal Hall coefficient were measured 
by a steady state technique. 
One edge of the crystal was glued to a cold head of 
a closed cycle refrigerator by varnish (GE7031), 
while a resistive heater ($\sim 120$ $\Omega $, KYOWA strain gauges) 
was attached on the opposite edge. 
Using the resistive heater, a longitudinal temperature gradient $\frac{dT_x}{dx}$ 
was generated. Through the measurement, the heat delivered to the hot end of the crystal 
was kept constant (1.32W).  
In the presence of the magnetic field (CRYOGENIC, MINI-CFM) 
applied perpendicular to 
$\frac{dT_x}{dx}$, the transversal temperature 
gradient $\frac{dT_y}{dy}$ was detected along the binary axis. (See Fig. 1.) 
Both gradients were evaluated from temperature differences measured 
with a pair of Au(Fe0.07\%)-Chromel differential thermocouples 
and the distances between the thermocouples $\bigtriangleup x=3.55$ mm and 
$\bigtriangleup y=2.0$ mm. Since the thermal conductivity of chromel is 13.8 W/mK, 
which is comparable to $\sim 10$ W/mK of bismuth, 
an error of the heat current estimation should be below 0.15\% using 
the chromel wire with 50 micron diameter and 2 cm length. 
To avoid leakage of the heat current by radiation and conduction through the air, 
we attached Cu radiation shield, and evacuated the air down to 10$^{-4}$ Pa. 
The loss of the heat current was estimated to be below 1\%.
We observed $\frac{dT_y}{dy}\not=0$ at 0 T due to a small misalignment of 
the positions of the thermocouple. This component was carefully subtracted by 
varying the magnetic field between -3 and +3. 

\begin{figure}[t]
\begin{center}
\vspace*{0cm}
\includegraphics[width=6cm,clip]{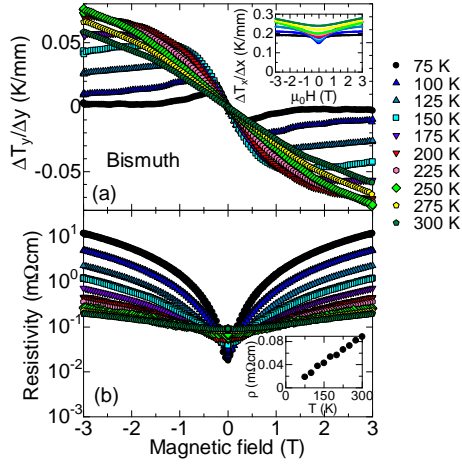}
\caption{(Color online) The magnetic field dependence of (a) 
$\frac{\bigtriangleup T_y}{\bigtriangleup y}$ and (b) the 
resistivity. The insets of Figs. 1(a) and (b) show the magnetic field dependence 
of $\frac{\bigtriangleup T_x}{\bigtriangleup x}$ and the temperature dependence 
of the resistivity under zero field, 
respectively. 
}
\end{center}
\end{figure}

Figure 1(a) shows the magnetic field dependence of the temperature 
gradient $\frac{\bigtriangleup T_y}{\bigtriangleup y}$. 
$\frac{\bigtriangleup T_y}{\bigtriangleup y}$ decreases with increasing the magnetic field, 
which indicates a negative sign of $R_{\rm TH}$. The 
magnitude of $\frac{\bigtriangleup T_y}{\bigtriangleup y}$ at 250 K and 
3 T is 0.076 K/mm, which is about 29 \% of 0.26 K/mm in 
$\frac{\bigtriangleup T_x}{\bigtriangleup x}$ shown in the inset of Fig. 1(a). 
Thus, $R_{\rm TH}$ is 
evaluated to be $-$0.1 $T^{-1}$ at 250 K and 3 T. 
This is consistent with $R_{\rm TH}$ in the reference \cite{smith}. 
We would like to point out that $R_{\rm TH}$ is $-$0.39 T$^{-1}$ at 0.33 T and 150 K, 
which is one of the largest $R_{\rm TH}$ values among metal elements and 
semiconductors. \cite{HgSe}. 
The nonlinear character of $\frac{\bigtriangleup T_y}{\bigtriangleup y}$ 
is caused by the changes of the physical properties under magnetic 
field as shown in Fig. 2. 
To determine $\kappa _{\rm el}$ and $\mu _{\rm H}$, 
resistivity $\rho $ was measured under magnetic field 
as shown in Fig. 1(b). 
With increasing the magnetic field, $\rho $ remarkably increases. This 
large positive magnetoresistance is one of the significant properties of 
the bismuth. The inset of Fig. 1(b) shows $\rho $($T$) under zero field. 
This is also a typical behavior (80 $\mu \Omega $cm at 300 K) of 
a bismuth single crystal.

\begin{figure}[t]
\begin{center}
\vspace*{0cm}
\includegraphics[width=6.5cm,clip]{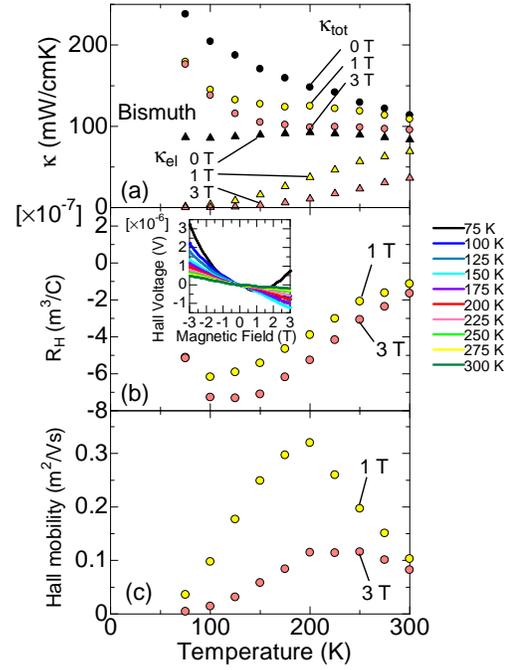}
\caption{(Color online) Temperature dependence of the thermal conductivity, (b) 
Hall coefficient, and (c) Hall mobility. The inset of Fig. 2(b) shows the magnetic 
field dependence of the Hall voltage. 
}
\end{center}
\end{figure}

Using the Fourier's law $j_x =-\kappa _{\rm tot}\frac{\bigtriangleup T_x}{\bigtriangleup x}$, 
the total thermal conductivity $\kappa _{\rm tot}$ 
was obtained as a function of temperature as shown in Fig. 2(a). 
$\kappa _{\rm tot}$ is a sum of the lattice thermal conductivity 
$\kappa _{\rm ph}$ and the electron thermal conductivity $\kappa _{\rm el}$ 
($\kappa _{\rm tot}=\kappa _{\rm ph} +\kappa _{\rm el}$). 
$\kappa _{\rm el}$ is evaluated by the Wiedemann-Franz law, 
$\kappa _{\rm el}=L_0 T\sigma $ ($L_0$: Sommerfeld's value $2.44\times 10^{-8}$ 
W$\Omega $/K$^2$, $\sigma$: electrical conductivity). 
At 300 K, the magnitude of $\kappa _{\rm tot}$ is 110 mW/cmK, 
which is close to 100 mW/cmK in the reference \cite{gallo}.
The lower thermal conductivity than that of a conventional metal such as 
Cu, Ag, or Au 
is attributed to both the relatively low electrical conductivity and low $\kappa _{\rm ph}$. 

Figure 2(b) shows the temperature dependence of the Hall coefficient $R_{\rm H}$. 
The sign of $R_{\rm H}$ is negative, which indicates 
that major carriers are electrons. This is consistent with the sign of $R_{\rm TH}$. 
The magnitude of $R_{\rm H}$ is 
$1\times 10^{-7}$ m$^3$/C at 300 K and 1 T, 
which shows a low carrier concentration $n$ of 6.2$\times $ $10^{19}$cm$^{-3}$.
With increasing magnetic field, $R_{\rm H}$ slightly increases. 
This result also reproduces $R_{\rm H}$ measured by Rosenbaum {\it et al.} \cite{rosenbaum} 
Using the relationship $\mu _{\rm H}=R_{\rm H}/\rho $, 
the Hall mobility $\mu _{\rm H}$ was derived as shown in figure 2(c). 
As expected, a rather large $\mu _{\rm H}$ was observed. 
At around 200 K, peak structure was appeared, in which decreased 
$\mu _{\rm H}$ at around room temperature is 
attributed to the increased phonon scatterings of conduction electrons \cite{zee}.

\begin{figure}[t]
\begin{center}
\vspace*{0cm}
\includegraphics[width=6.5cm,clip]{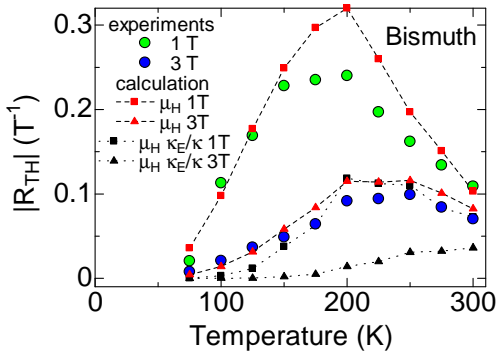}
\caption{(Color online) Temperature dependence of thermal Hall coefficient at 1 and 3 T. 
The dotted lines represent theoretical calculation using a single band model of the thermal 
Hall coefficient. 
}
\end{center}
\end{figure}

Now we compare $R_{\rm TH}$($T$) of bismuth with Eqs. (1) and (2). 
The measured $\kappa _{\rm el}$, $\kappa _{\rm tot}$, 
and $\mu _{\rm H}$ in Fig. 2 were used for the calculations. 
Figure 3 shows $R_{\rm TH}$ ($T$) at 1 and 3 T. 
The dotted and broken lines represent the calculations. 
We found that $\mu _{\rm H}$ well reproduces $R_{\rm TH}$($T$), 
which is consistent with the previous 
results that the conventional metals obey Eq. (1). 
On the other hand, Eq. (2) that describes $R_{\rm TH}$($T$) of the semiconductor 
does not give a reasonable fitting of $R_{\rm TH}$($T$). 
The reason why heat conduction throuth lattices does not decrease 
$R_{\rm TH}$ of bismuth is an open question. 
Here, we briefly discuss a two-band model. 
As a first order approximation (magnetic field $B\sim 0$), 
the thermal Hall coefficient $R'_{\rm TH}$ for a two band model is described as 
$\kappa R'_{\rm TH}=\kappa _p R_{{\rm TH}p}+\kappa _n R_{{\rm TH}n}$ 
$=\kappa _p \mu _{{\rm H}p}-\kappa _n \mu _{{\rm H}n}$ \cite{wilson}, 
where $p$ and $n$ indexes represent a hole- and electron-like band, respectively. 
In the referece \cite{rosenbaum}, a ratio of hole and electron concentrations 
$c=\frac{p}{n}=0.55$, electron mobility $\mu _{{\rm H}n}=0.43$ m$^2$V$^{-1}$s$^{-1}$, and 
hole mobility $\mu _{{\rm H}p}=0.15$ m$^2$V$^{-1}$s$^{-1}$ at 270 K in a bismuth thin film were reported. 
If one applies these parameter to the equation shown above, 
$R'_{\rm TH}$ is evaluated to be 93 \% of $R_{{\rm TH}n}$. 
This implies that electron transport is dominant in bismuth. 
Concerning the deviation from Eq. (2), there are two possibile origins; 
(1) Dirac fermion with a linear dispersion, and (2) interband scattering. 
Bismuth is known as a Dirac fermion sysytem \cite{ong}, where 
linear-dispersion relation is realized. This relation gives $m^*\cong 0$, 
which can enlarge $R_{\rm TH}$. Concerning the possibility (2), 
the two band model shown above does not take into account the interaction 
between the electron band and the hole band (i.e. interband scattering) 
that explains large diamagnetism in bismuth \cite{fukuyama}. This effect may 
increase $R_{\rm TH}$. 
As a further analysis, a theoretical calculation of $R_{\rm TH}$ on bismuth 
based on two-bands Dirac fermion taking into account 
interband scatterings will be desired.

Lastly, we would like to add a few comments on an improvement of the 
magnitude of $R_{\rm H}$. 
According to the Eqs. (1) and (2), small $\kappa _{\rm ph}$, large $\kappa _{\rm el}$, and 
large $\mu _{\rm H}$ are essential for large $R_{\rm TH}$. 
$\mu _{\rm H}$ is described as $\mu _{\rm H}=\frac{e \tau }{m^*(n)}$, 
where $e$, $\tau $, and $m^*$ represent electron charge, electron scattering time, and 
effective mass of electron, respectively. 
With increasing $n$, 
$\kappa _{\rm el}$ increases while $\mu _{\rm H}$ decreases due to increase of 
$m^*$, which gives an optimum value of $R_{\rm TH}$ at 
an optimum $n$. To achieve a further enhancement of $R_{\rm TH}$, 
one needs to control $\tau $. 
Recent theoretical study on thermoelectric performance in 
bismuth thin film reveals a significant increase of the 
thermoelectric properties mainly due to the enhanced electron 
scattering time in the Bi thin film with $\sim 10$-nm thickness \cite{takahashi}, 
in which quantum spin Hall state appears due to the 
strong spin-orbit interaction. To keep a sign of the spin momentum in the QSH state, 
the electron is not scattered by impurities, which realizes an increase of $\mu _{\rm H}$. 
In the same manner, $R_{\rm TH}$ can largely enhance 
when $\mu _{\rm H}$ increases with decreasing a thickness of a bismuth thin film. 
Again, we would like to emphasize that Dirac fermion with linear-dispersion relation will give 
$m^*\cong 0$, which also enlarges $R_{\rm TH}$. 
Thus, topological insulators and Dirac fermion systems such as 
Bi$_2$Se$_3$, Bi$_2$Te$_3$ or graphene can be a good 
playground for finding a large $R_{\rm TH}$. 
The large $R_{\rm TH}$ under low magnetic field in bismuth 
will open a route for controlling heat current in electronic devices. 

In conclusion, we have measured resistivity, Hall coefficient, 
thermal Hall coefficient, and thermal conductivity of a bismuth 
single crystal and found that the large thermal Hall coefficient is 
well explained by a theoretical calculation. 
A high Hall mobility and a low thermal conductivity can cause 
the large thermal Hall coefficient in bismuth. 
In particular, $\sim $29\% of the heat current along the bisectrix direction 
is found to be oriented along the binary axis at 250 K and 3 T. 
We also have discussed that the strong spin-orbit interaction can 
be a driving force of a further enhancement of the thermal Hall coefficient.

We would like to thank S. Murakami for fruitful discussion on the thermoelectric properties 
of Bi thin film, and also thank D. Sawaki for technical support on the thermal conductivity 
measurement.

\end{document}